\documentclass[aps,showpacs,twocolumn,draft]{revtex4}

\usepackage{latexsym}

\begin{document}


\title{Hyperbolic slicings of spacetime: singularity avoidance and
gauge shocks}

\author{Miguel Alcubierre}

\affiliation{Instituto de Ciencias Nucleares, Universidad Nacional
Aut\'onoma de M\'exico, A.P. 70-543, M\'exico D.F. 04510, M\'exico.}

\date{October, 2002}


\begin{abstract}
I study the Bona-Masso family of hyperbolic slicing conditions,
considering in particular its properties when approaching two
different types of singularities: focusing singularities and gauge
shocks.  For focusing singularities, I extend the original analysis of
Bona {\em et. al.} and show that both marginal and strong singularity
avoidance can be obtained for certain types of behavior of the slicing
condition as the lapse approaches zero.  For the case of gauge
shocks, I re-derive a condition found previously that eliminates them.
Unfortunately, such a condition limits considerably the type of
slicings allowed.  However, useful slicing conditions can still be
found if one asks for this condition to be satisfied only
approximately.  Such less restrictive conditions include a particular
member of the 1+log family, which in the past has been found
empirically to be extremely robust for both Brill wave and black hole
simulations.
\end{abstract}


\pacs{
04.20.Ex, 
04.25.Dm, 
95.30.Sf, 
}


\maketitle


\section{Introduction}
\label{sec:introduction}

The choice of good coordinates plays a crucial role in finding
solutions of the Einstein equations.  This is particularly important
in the case of numerical simulations of strongly gravitating systems,
where a bad coordinate choice can easily lead to the formation of a
coordinate singularity which stops the numerical simulation and
severely limits the region of the spacetime covered by it.  Coordinate
singularities are not the only concern: the presence of physical
singularities, such as those associated with black holes, can also
have a deep impact in a numerical simulation, as unless special
care is taken the time slices can march right onto the physical
singularity very early during a simulation.

When considering dynamical evolutions of spacetime based on a 3+1
decomposition~\cite{Arnowitt62,York79}, the coordinate choice
naturally separates in two different aspects: the choice of a specific
foliation of spacetime into spatial hyper-surfaces (also known as the
``slicing''), associated with the lapse function $\alpha$, and the
choice of the way in which the lines of constant spatial coordinates
(the ``time lines'') propagate from one hyper-surface to the next,
associated with the shift vector $\beta^i$.  Here I will concentrate
fully in the role played by the choice of a slicing condition and
leave the choice of a shift vector for a later work.

In order to specify a foliation of spacetime, one needs to prescribe a
way to calculate the lapse function $\alpha$, which measures the
proper time interval between neighboring hyper-surfaces along their
normal direction.  There is, of course, an infinite number of ways in
which one can choose the lapse function, but typically the different
choices can be classified in the following way: 1) Prescribed
slicings, where the lapse is specified as an {\em a priori} known
function of space and time, 2) algebraic slicing conditions, where the
lapse is specified algebraically as some function of the geometric
variables (metric and extrinsic curvature) at each hyper-surface, 3)
elliptic slicing conditions, where the lapse is obtained by solving an
elliptic differential equation at every time step that typically
enforces some geometric condition on the spatial hyper-surfaces, and
4) time derivative slicing conditions, where the time derivative of
the lapse is specified as some algebraic function of the geometric
quantities and the lapse is evolved as just another dynamical quantity
(this last case is often included in the algebraic slicing class
mentioned above).

An example of a prescribed slicing is the so-called ``geodesic
slicing'', where one simply takes $\alpha=1$.  An example of an
algebraic slicing condition is ``harmonic slicing'' where one takes
instead $\alpha=\sqrt{\gamma}$, with $\gamma$ the determinant of the
spatial metric.  A well known elliptic slicing condition is the
``maximal slicing'' condition~\cite{Smarr78b}, which requires that the
spatial volume elements remain constant during the evolution.
Elliptic conditions are typically robust and well behaved, but have
the drawback of being computationally expensive.  Algebraic conditions
are much easier to apply, but are difficult to analyze in a general
case.  Time derivative slicing conditions, on the other hand, have the
advantage of being both easy to implement and, in the particular case
when they lead to hyperbolic equations (as is the case of the Bona-Masso
family~\cite{Bona94b}, see below), much easier to analyze as
well. They also include many well known algebraic conditions as their
integral form.

In this paper, I will only consider hyperbolic slicing conditions and
study in which ways such conditions can lead to pathological slicings
and how can those pathologies best be avoided.  There are many
different ways in which a foliation of spacetime can become
pathological: The slices can hit a physical singularity, the slices
can hit a coordinate singularity where the volume elements become zero
(the normal lines focus), the slices can become non-smooth at some
point or the slices can remain smooth but stop being space-like (they
can become null at a point, for example).  Of the different possible
pathologies mentioned above, I will concentrate on two specific types:
``focusing singularities''~\cite{Bona88} defined as those for which
the spatial volume elements vanish at a bounded rate, and ``gauge
shocks''~\cite{Alcubierre97a} defined as solutions for which the lapse
becomes discontinuous as a consequence of the crossing of the
characteristic lines associated with the propagation of the gauge, and
the time slices therefore become non-smooth.

This paper is organized as follows: In Sec.~\ref{sec:BonaMasso}, I
introduce the Bona-Masso hyperbolic slicing condition.  Focusing
singularities are defined in Section~\ref{sec:focusing} where I also
find under which circumstances the Bona-Masso slicing condition is
singularity avoiding.  In Sec.~\ref{sec:shocks}, I introduce the idea of a
gauge shock, derive a condition to avoid them and see what slicings
obey that condition either exactly or approximately.  I conclude in
Sec.~\ref{sec:conclusion}.


\section{The Bona-Masso family of hyperbolic slicing conditions}
\label{sec:BonaMasso}

The Bona-Masso family of slicing conditions~\cite{Bona94b} is a time
evolution type of condition for which the lapse is chosen to satisfy
the following evolution equation
\begin{equation}
\frac{d}{dt} \; \alpha \equiv \left( \partial_t - {\cal L}_\beta \right)
\alpha = - \alpha^2 f(\alpha) \, K \\ ,
\label{eq:BonaMasso}
\end{equation}
with ${\cal L}_\beta$ the Lie derivative with respect to the shift
vector $\beta^i$, $K$ the trace of the extrinsic curvature and
$f(\alpha)$ a positive but otherwise arbitrary function of $\alpha$.
The reason why $f(\alpha)$ has to be positive will become clear below.
Here we just mention the fact that the right hand side of
condition~(\ref{eq:BonaMasso}) is the most general term one can
construct that involves only first order spatial scalars~\cite{Bona94b}.

There are several things that are important to mention about the
family of slicing conditions~(\ref{eq:BonaMasso}).  First, we notice
that even if this family was proposed in the context of the Bona-Masso
hyperbolic reformulation of the Einstein
equations~\cite{Bona89,Bona92,Bona93,Bona94b,Bona97a}, it is in fact
quite general and can be used successfully with any particular form of
the 3+1 evolution equations.  This was shown recently when such a
condition was used together with the Baumgarte-Shapiro
Shibata-Nakamura (BSSN) formulation~\cite{Baumgarte99,Shibata95} to
obtain long-term stable and accurate evolutions of black hole
spacetimes~\cite{Alcubierre01a,Alcubierre02a}.  Also,
condition~(\ref{eq:BonaMasso}) is a generalization of slicing
conditions that have been used in evolution codes based on the
Arnowitt-Deser-Misner (ADM) formulation~\cite{Arnowitt62,York79} since
the early 90's~\cite{Bernstein93a,Anninos94c}.

Second, condition~(\ref{eq:BonaMasso}) can also be trivially adapted
to the case when, instead of the lapse $\alpha$, one evolves a
densitized lapse of the form
\begin{equation}
Q := \alpha \, \gamma^{\sigma/2}
\end{equation}
with $\gamma$ the determinant of the spatial metric $\gamma_{ij}$ and
$\sigma$ a constant parameter.  Such a densitized lapse (particularly
the case $\sigma=-1$) has recently been advocated in the context of
hyperbolic reformulations of the Einstein equations (see for
example~\cite{Anderson99,Kidder01a}).  In terms of $Q$,
condition~(\ref{eq:BonaMasso}) becomes
\begin{equation}
\frac{d}{dt} \; Q := - \frac{Q^2}{\gamma^{\sigma/2}} \left( f + \sigma
\right) K \\ ,
\end{equation}
where to calculate the Lie derivative with respect to $\beta^i$
contained in the operator $d/dt$ one must use the fact that $Q$ is a
density of weight $\sigma$.

Finally, it is also important to mention that the shift terms included
through the Lie derivative in condition~(\ref{eq:BonaMasso}) are such
that one is guaranteed to obtain precisely the same spacetime
foliation regardless of the value of the shift vector.  This would
seem to be a natural requirement for any slicing condition. However,
it is plausible that in a particular situation one would like to
choose a slicing condition and a shift vector that are closely
interrelated (see for example~\cite{Balakrishna96a}).  Indeed,
generalizations of the Bona-Masso slicing condition that in the
presence of a non-zero shift vector do not have the
form~(\ref{eq:BonaMasso}) have already been used in the literature.
For example, Refs.~\cite{Alcubierre00a,Yo02a} use \mbox{$\partial_t
\alpha = - \alpha f ( \alpha K - D_i \beta^i)$} instead
of~(\ref{eq:BonaMasso}) (with $D_i$ the spatial covariant derivative),
and Ref.~\cite{Alcubierre02a} simply uses \mbox{$\partial_t \alpha = -
\alpha^2 f (K- K_0)$} (with $K_0 = K(t=0))$ even with a non-zero shift
vector.  Which off these generalizations is best under different
circumstances is an important question that I will leave for a future
work.

\vspace{5mm}

With these comments in mind, let us go back to
condition~(\ref{eq:BonaMasso}).  Taking an extra time derivative we
find
\begin{equation}
\frac{d^2}{dt^2} \; \alpha = - \alpha^2 f \left[ \frac{d}{dt} K
- \alpha ( 2 f + \alpha f' ) K^2 \right] ,
\end{equation}
with $f':= df/d\alpha$.  From the ADM evolution equations one easily
finds that, in vacuum,
\begin{equation}
\frac{d}{dt} \; K = - D^2 \alpha + \alpha K_{ij} K^{ij} \, ,
\end{equation}
with $D^2$ the Laplace operator associated with the spatial metric.
This last equation implies
\begin{eqnarray}
&& \frac{d^2}{dt^2} \; \alpha - \alpha^2 f D^2 \alpha
= \nonumber \\
&& \hspace{3em} - \alpha^3 f \left[ K_{ij} K^{ij} - \left( 2 f
+ \alpha f' \right) K^2 \right] .
\label{eq:lapsewave}
\end{eqnarray}

Equation~(\ref{eq:lapsewave}) shows that the lapse obeys a wave
equation with a quadratic source term in $K_{ij}$.  It is because of
this that we say the slicing condition~(\ref{eq:BonaMasso}) is a
hyperbolic slicing condition: it implies that the lapse evolves with a
hyperbolic equation (but see the discussion on gauge shocks below for
a more formal proof of hyperbolicity).  The wave speed associated with
equation~(\ref{eq:lapsewave}) along a specific direction $x^i$ can be
easily seen to be
\begin{equation}
v_g = \alpha \sqrt{f \gamma^{ii}} \, .
\label{eq:gaugespeed}
\end{equation}

Notice that this will only be real if $f(\alpha) \geq 0$, which
explains why we asked for $f(\alpha)$ to be positive.  In fact,
$f(\alpha)$ must be strictly positive because if it was zero there
would be no complete set of eigenvectors and we would not have a
strongly hyperbolic system (see Sec.~\ref{sec:gaugeshocks}).

To see how the gauge speed $v_g$ is related to the speed of light
consider for a moment a null world-line.  It is not difficult to find
that such a world-line will have a coordinate speed along direction
$x^i$ given by
\begin{equation}
v_l = \alpha \sqrt{\gamma^{ii}} \, ,
\end{equation}
so the gauge speed~(\ref{eq:gaugespeed}) can be smaller or larger that
the speed of light depending on the value of $f$.

Notice that having a gauge speed that is larger than the speed of
light does not lead to any causality violations, as the superluminal
speed is only related with the propagation of the coordinate system.
One could argue that superluminal gauge speeds are not desirable as
they would allow gauge effects to propagate out of black hole
horizons, for example.  Empirically, however, the most successful
slicing conditions for the simulation of black hole spacetimes have
been precisely those that have superluminal (even extremely large)
gauge speeds: maximal slicing, which as an elliptic condition has, at
least formally, an infinite speed of propagation, and the 1+log
slicing condition which is a member of the Bona-Masso family that has
superluminal gauge speeds whenever the lapse is small (see following
section).


\subsection{Relating lapse and spatial volume elements}
\label{sec:relating}

The ADM evolution equation for the spatial metric $\gamma_{ij}$ is
given by
\begin{equation}
\frac{d}{dt} \; \gamma_{ij} = - 2 \alpha K_{ij} \, ,
\end{equation}
which implies the following evolution equation for the spatial
volume elements $\gamma^{1/2}$:
\begin{equation}
\frac{d}{dt} \; \gamma^{1/2} = - \alpha \gamma^{1/2} K \, .
\label{eq:sqrtgammadot}
\end{equation}

Comparing this with equation~(\ref{eq:BonaMasso}), and taking
$f=1$, we can trivially solve for $\alpha$ in terms of $\gamma^{1/2}$
to obtain (in the case of zero shift vector)
\begin{equation}
\alpha = h(x^i) \; \gamma^{1/2} \, ,
\label{eq:harmonic}
\end{equation}
with $h(x^i)$ a time independent function.  It is very important to
stress the fact that the previous relation holds only when moving
along the normal direction to the hypersurfaces, and not when moving
along the time lines which will differ from the normal direction for
any non-zero shift vector.  That is, we are relating the lapse to the
volume elements as seen by the normal observers.  In the following,
whenever we relate the lapse to the volume elements, it should always
be understood that we are referring to the volume elements associated
with the normal observers.

It is not difficult to show (see following section) that
equation~(\ref{eq:harmonic}) is equivalent to the condition
\begin{equation}
\Box t = g^{\mu \nu} \Gamma^0_{\mu \nu} = 0 \, ,
\label{eq:boxt}
\end{equation}
with $g_{\mu \nu}$ the spacetime metric.  That is, $f=1$ corresponds
to the case when the time coordinate is a harmonic function.  Because
of this the case $f=1$ is known as ``harmonic slicing''.  Notice also
that in this case the gauge speed is identical to the physical speed
of light, {\em i.e.} the gauge propagates along null lines.  Harmonic
slicing can be seen to be equivalent to having a time independent
lapse density $Q$ of weight $\sigma=-1$ (again, assuming the shift
vanishes).

One can construct other well known families of slicing conditions by
choosing different forms of $f(\alpha)$.  For example, if we choose
$f(\alpha) = N$, with $N$ a constant, we obtain what can be called the
``generalized harmonic slicing condition'', which can also be easily
integrated to give
\begin{equation}
\alpha = h(x^i) \, \gamma^{N/2} \, .
\end{equation}

And if we take $f(\alpha)=N/\alpha$ we obtain the ``1+log''
family~\cite{Abrahams92a,Anninos93b}, which again can be
integrated to find
\begin{equation}
\alpha = h(x^i) + \ln \left( \gamma^{N/2} \right) \, .
\end{equation}

The 1+log family mimics maximal slicing and has strong singularity
avoiding properties (see section~\ref{sec:focusing} below).  In
particular, 1+log slicing with $N=2$ has been found empirically to be
very robust when evolving black hole
spacetimes~\cite{Arbona99,Alcubierre01a,Alcubierre02a}.  As mentioned
before, one can easily see that the gauge speed associated with the
1+log family can become far larger than the speed of light as the
lapse becomes smaller ($v_g/v_l = \sqrt{N/\alpha}$).

More generally, using equation~(\ref{eq:sqrtgammadot}) one can find
that for arbitrary $f(\alpha)$ the following relation between $\alpha$
and $\gamma^{1/2}$ holds
\begin{equation}
d \ln \gamma^{1/2} = \frac{d \alpha}{\alpha f(\alpha)} \; ,
\label{eq:dtgamma_dtalpha}
\end{equation}
which implies
\begin{equation}
\gamma^{1/2} = F(x^i) \, \exp \left\{ \int{\frac{d \alpha}
{\alpha f(\alpha)} } \right\} \; .
\label{eq:gammaofalpha}
\end{equation}
with $F(x^i)$ again a time independent function.  This last expression
will be the starting point when we discuss focusing singularities
below.


\subsection{The foliation equation}
\label{sec:foliation}

Consider now a spacetime with metric $g_{\mu \nu}$, and assume that we
have a foliation of this spacetime into spatial hypersurfaces.  Such a
foliation allows us to define a time function $T$ whose level sets
correspond to the members of the foliation. One can show that the
Bona-Masso slicing condition~(\ref{eq:BonaMasso}) can be written as a
generalized wave equation for the time function $T$ in the following
way
\begin{equation}
\left[ g^{\mu \nu} + \left( 1 - \frac{1}{f(\alpha)} \right)
n^\mu n^\nu \right] T_{;\mu \nu}  = 0 \, ,
\label{eq:foliation}
\end{equation}
with $n^\mu$ the unit normal vector to the spatial hypersurfaces.  I
will call equation~(\ref{eq:foliation}) the ``foliation equation''.
Notice that for harmonic slicing the above equation reduces to the
simple wave equation~(\ref{eq:boxt}).

The fact that the foliation equation above is equivalent to the
Bona-Masso slicing condition~(\ref{eq:BonaMasso}) can be shown by
choosing 3+1 coordinates $\{t,x^i\}$ adapted to the foliation.  In
this coordinate system we can take $T=t$, so
equation~(\ref{eq:foliation}) becomes
\begin{equation}
\left[ g^{\mu \nu} + \left( 1 - \frac{1}{f(\alpha)} \right) n^\mu n^\nu
\right] \Gamma^0_{\mu \nu} = 0 \; .
\label{eq:foliation2}
\end{equation}
Using now the 3+1 expressions for the components of the 4-metric
$g_{\mu \nu}$ we obtain the following expressions for the Christoffel
symbols
\begin{eqnarray}
\Gamma^0_{00} &=& \frac{1}{\alpha} \left( \partial_t \alpha
+ \beta^i \partial_i \alpha - \beta^i \beta^j K_{ij} \right) \; , \\
\Gamma^0_{0i} &=& \frac{1}{\alpha} \left( \partial_i \alpha
- \beta^m K_{im} \right) \; , \\
\Gamma^0_{ij} &=& - \frac{1}{\alpha} K_{ij} \; .
\end{eqnarray}
Substituting this into~(\ref{eq:foliation2}), and using the 3+1
expressions for the normal vector $n^\mu$, we find
\begin{equation}
\partial_t \alpha - \beta^i \partial_i \alpha + \alpha^2 f(\alpha) K = 0
\; ,
\end{equation}
which is precisely the Bona-Masso condition~(\ref{eq:BonaMasso}).

Notice that if we take $f>1$, one can always have a unit normal vector
$n^\mu$ such that the coefficient of the $\partial^2_t T$ term in the
foliation equation changes sign, and the equation apparently becomes
elliptic.  The system is in fact still hyperbolic (see
Sec.~\ref{sec:gaugeshocks}), and the change in signature just reflects
the fact that for $f>1$ the characteristic cones can tilt beyond the
time axis.

The foliation equation~(\ref{eq:foliation}) will prove to be very
important when we study the formation of gauge shocks.


\section{Focusing singularities}
\label{sec:focusing}

A very important property of slicing conditions is that of
``singularity avoidance''.  Singularity avoidance refers to the
property of certain slicing conditions of slowing down coordinate
time, by making the lapse go to zero, when the spatial volume elements
$\sqrt{\gamma}$ go to zero (this is known as the ``collapse of the
lapse'').  Recent advances in black hole excision
techniques~\cite{Thornburg87,Thornburg93,Seidel92a,Anninos94e} would
seem to minimize the need for singularity avoidance in the choice of
slicing conditions.  One should remember, however, that singularity
avoidance is not only important when one is interested in studying
black hole spacetimes where real physical singularities are present,
but is also needed in order to prevent the formation of coordinate
singularities caused by the focusing of the normal lines in regions
with strong gravitational fields.

Bona {\em et. al.} have shown~\cite{Bona97a} that the slicing
condition~(\ref{eq:BonaMasso}) can avoid so-called ``focusing
singularities'' for some choices of the function $f(\alpha)$.  Here I
will extend their analysis and show explicitly what type of behavior
$f(\alpha)$ must have as $\alpha$ approaches zero in order to avoid
such singularities.

Following~\cite{Bona97a}, we define a focusing singularity as a place
where the spatial volume elements $\gamma^{1/2}$ vanish at a bounded
rate.  Let us assume that such a singularity occurs after a finite
proper time $\tau_s$ away from our initial time slice (as measured by
the normal observers).  From the definition of the lapse we see that
the elapsed coordinate time will then be
\begin{equation}
\Delta t = \int_0^{\tau_s} \frac{d \tau}{\alpha} \; .
\label{eq:deltat}
\end{equation}

We will further characterize the singularity by the rate at which
$\gamma^{1/2}$ approaches zero as a function of proper time.  We will
say a singularity is of order $m$ if $\gamma^{1/2}$ approaches zero as
\begin{equation}
\gamma^{1/2} \sim \left( \tau_s - \tau \right)^m \, ,
\label{eq:orderm}
\end{equation}
with $m$ some constant power.  Notice that $m$ must be strictly
positive for there to be a singularity at all, and it must be larger
than or equal to $1$ for the singularity to be approached at a bounded
rate.

As the volume elements $\gamma^{1/2}$ approach zero, there are clearly
three possible behaviors for the lapse: 1) $\alpha$ remains finite as
$\gamma^{1/2}$ vanishes, 2) $\alpha$ vanishes as $\gamma^{1/2}$
vanishes, and 3) $\alpha$ vanishes before $\gamma^{1/2}$ vanishes.
Case 1 would clearly imply that coordinate time remains finite at the
singularity, so the singularity would {\em not} be avoided.  However,
from equation~(\ref{eq:gammaofalpha}) one can easily see that if the
lapse remains always finite it is impossible for the volume elements
to ever vanish (remember that $f(\alpha)$ is never allowed to be
zero).  We then conclude that case 1 can never happen, which implies
that the Bona-Masso slicing condition~(\ref{eq:BonaMasso}) always
causes the lapse to collapse when the volume elements approach zero,
for any $f(\alpha)>0$.  Case~3, on the other hand, implies that the
time slices stop advancing a finite coordinate time before the
singularity is reached (the time slices can in fact move back under
certain conditions, see below).  We will call such behavior ``strong
singularity avoidance''.  Finally, case 2 corresponds to the case when
the lapse becomes zero at the same time as the volume elements.
Whether in such a case the singularity is reached after a finite or
infinite coordinate time will depend on the speed at which $\alpha$
approaches zero at the singularity.  We will say that a slicing is
``marginally singularity avoiding'' if the singularity is reached
after an infinite coordinate time.

To study under which conditions we can have strong or marginal
singularity avoidance we must say something about the form of the
function $f(\alpha)$.  From now on we will therefore assume that, as
$\alpha$ approaches zero, the function $f(\alpha)$ behaves as
\begin{equation}
f(\alpha) = A \alpha^n \; ,
\label{eq:f(alpha)}
\end{equation}
with both $A$ and $n$ constants and $A>0$.  Such an assumption implies
that
\begin{eqnarray}
\int{\frac{d \alpha}{\alpha f(\alpha)}} &=& \frac{1}{A}
\int{\frac{d \alpha}{\alpha^{n+1}}} \nonumber \\
&=& \left\{
\begin{array}{ll}
\ln \alpha^{1/A}   & \quad n = 0 \\
-1/(n A \alpha^n)  & \quad n \neq 0
\end{array} \right. 
\end{eqnarray}

Let us first consider the case $n \neq 0$.
From equation~(\ref{eq:gammaofalpha}) we now find that
\begin{equation}
\gamma^{1/2} \sim \exp \left( - \frac{1}{n A \alpha^n} \right) \; .
\end{equation}
As $\alpha$ approaches zero we have two separate cases depending on
the sign of $n$:
\begin{equation}
\lim_{\alpha \rightarrow 0} \gamma^{1/2} = \left\{
\begin{array}{ll}
\mbox{finite}  & \quad n < 0 \\
0              & \quad n > 0
\end{array} \right. 
\end{equation}
Since for $n<0$ the volume elements remain finite as the lapse
approaches zero we conclude that such a case corresponds to strong
singularity avoidance.  On the other hand, for $n>0$ both the lapse
and the volume elements go to zero at the same time so we can at most
have marginal singularity avoidance.

For the case $n=0$ we find, again using~(\ref{eq:gammaofalpha}), that
\begin{eqnarray}
&& \gamma^{1/2} \sim \alpha^{1/A} \nonumber \\
&& \Rightarrow \alpha \sim \gamma^{A/2} \; .
\end{eqnarray}
It is then clear that in this case $\alpha$ and $\gamma^{1/2}$ also
vanish at the same time.

We now need to decide if the cases $n>0$ and \mbox{$n=0$}, for which
the lapse and the volume elements become zero at the same time, reach
the singularity in an infinite or a finite coordinate time.  For this
we need to study the behavior of $\alpha$ as a function of proper time
$\tau$ as we approach the singularity.  Starting from
equation~(\ref{eq:deltat}) for the elapsed coordinate time we find
\begin{eqnarray}
\Delta t &=& \int_0^{\tau_s} \frac{d \tau}{\alpha} \nonumber \\
&=& \int_{\alpha_0}^{0} \frac{d \tau / d \alpha}{\alpha}
\; d \alpha \; ,
\label{eq:deltat2}
\end{eqnarray}
where $\alpha_0$ is the initial lapse and where we are already
assuming that we are interested in the case when $\alpha$ vanishes at
$\tau_s$.  Equation~(\ref{eq:deltat2}) implies that if $d \tau / d
\alpha$ remains different from zero as we approach the singularity
then $\Delta t$ will diverge and we will have marginal singularity
avoidance.  On the other hand, if $d \tau / d \alpha$ vanishes at the
singularity as $\alpha^p$ (with $p$ some positive power) or faster,
then the integral will converge and the singularity will be reached in
a finite coordinate time.

To find the behavior of $d \tau / d \alpha$ as we approach the
singularity, we notice that equation~(\ref{eq:orderm}) implies
\begin{equation}
d \gamma^{1/2} / d \tau \sim
- m \left( \tau_s - \tau \right)^{m-1} \; ,
\end{equation}
and
\begin{equation}
\frac{d \ln \gamma^{1/2}}{d \tau} = 
- \frac{m} {\left( \tau_s - \tau \right)}  \; .
\end{equation}
From this, together with equation~(\ref{eq:dtgamma_dtalpha}), we now find
\begin{equation}
\frac{d \alpha / d \tau}{\alpha f(\alpha)} =
- \frac{m} {\left( \tau_s - \tau \right)}  \; ,
\end{equation}
which can be integrated to give
\begin{equation}
\tau = \tau_s - \exp \left( \frac{1}{m}
\int{\frac{d \alpha}{\alpha f(\alpha)}}
\right) \; .
\end{equation}

If we now take $f(\alpha)$ given by~(\ref{eq:f(alpha)}) we finally
find
\begin{eqnarray}
\tau &=& \tau_s - \exp \left( \frac{1}{m A}
\int{\frac{d \alpha}{m \alpha^{n+1}}} \right) \nonumber \\
&=& \left\{
\begin{array}{ll}
\tau_s - \alpha^{1/mA}  & \quad n = 0 \\
\tau_s - \exp \left[ - 1 / (n m A \alpha^n) \right] & \quad n > 0
\end{array} \right.
\end{eqnarray}
The case $n<0$ is not of interest here since we already showed that for
such a case the lapse will vanish before we reach the singularity.

Let us consider the case $n>0$ first.  The derivative of $\tau$ with
respect to $\alpha$ then turns out to be
\begin{equation}
\frac{d \tau}{d \alpha} = - \frac{1}{m A \alpha^{n+1}} \;
\exp \left( - \frac{1}{n m A \alpha^n} \right) \; , 
\end{equation}
from which it is easy to see that as $\alpha$ approaches zero, $d \tau
/ d \alpha$ also approaches zero faster than any power.  As we have
seen, this means that the singularity is reached in a finite
coordinate time, so the case $n>0$ does not avoid the singularity (not
even marginally).

For $n=0$ we  have, on the other hand
\begin{equation}
\frac{d \tau}{d \alpha} = - \frac{1}{m A} \;
\alpha^{1/mA - 1} \; .
\end{equation}

We then see that
\begin{equation}
\lim_{\alpha \rightarrow 0} \frac{d \tau}{d \alpha} =
\left\{ \begin{array}{cl}
 0       & \quad m A < 1 \\
-1       & \quad m A = 1 \\
- \infty & \quad m A > 1
\end{array} \right.
\end{equation}

The case $m A < 1$ therefore reaches the singularity in a finite
coordinate time, while the cases $m A \geq 1$ reach it in an infinite
coordinate time and are therefore marginally singularity avoiding.

Our final result can be summarized as follows: If $f(\alpha)$ behaves
as~$f=A \alpha^n$ for small $\alpha$ and we have a singularity of
order $m$, then

\begin{enumerate}

\item For $n<0$ we have strong singularity avoidance.

\item For $n=0$ and $m A \geq 1$ we have marginal singularity avoidance.

\item For both $n>0$, and $n=0$ with $m A < 1$, we do not have
singularity avoidance, even though the lapse collapses to zero at the
singularity.

\end{enumerate}

In the particular case when we have a singularity of order $m=1$, then
harmonic slicing ($n=0$, $A=1$) marks the boundary between avoiding
and reaching the singularity.

As a final observation, let us consider again the case \mbox{$n<0$},
for which we have shown that we have strong singularity avoidance.
Looking at our original slicing condition~(\ref{eq:BonaMasso}) we see
that if $n \leq -2$ then, as the lapse approaches zero, one can not
guarantee that $\partial_t \alpha$ will also approach zero.  The lapse
can therefore easily become negative and the slices will not only
avoid the singularity but can in fact back away from it.  This type
of behavior is probably not desirable, as one runs the risk of having
the time slices stop being space-like (they advance in one region and
move back in another).  If we want to guarantee that we have strong
singularity avoidance without the lapse becoming negative we must
limit ourselves to the region $-2 < n < 0$.  Notice that the 1+log
family corresponds to $n=-1$, and is precisely in the middle of this
range, which probably accounts for the fact that empirically it has
been found to be a very good choice.


\section{Gauge shocks}
\label{sec:shocks}

In physics, one talks about ``shock waves'' as solutions to the
hydrodynamic equations where very sharp density gradients propagate
through a medium at speeds that are higher than the speed of sound in
that medium.  Mathematically, shocks are discontinuous solutions of a
non-linear hyperbolic system of equations characterized by the fact
that characteristic lines converge at the discontinuity.  The
discontinuity propagates with a speed called the ``shock speed'' that
is somewhere in between the values of the characteristic speeds in the
regions behind and in front of the shock.  Usually, in order to
completely determine the form and speed of a shock one needs to
supplement the evolution equations with extra conditions coming from
physical considerations known as ``entropy-conditions''(see for
example~\cite{Lax73}). Such entropy conditions are necessary
because once a discontinuity forms, the possible mathematical
extensions through it are no longer unique and one needs a mechanism
to choose the physically allowed solutions.

It is well known that physical shocks (i.e. shocks in the geometry) do
not appear in general relativity.  Solutions of the Einstein equations
normally called ``shock fronts'' refer to discontinuities in the
curvature of spacetime present in the initial data that propagate with
the speed of light.  These type of solutions are not shocks in a
mathematical sense but are called instead ``contact discontinuities''.
Here we will not consider discontinuities in the geometry, but rather
solutions to our hyperbolic slicing conditions that start from smooth
initial data and develop discontinuities later, when the
characteristic lines associated with the gauge cross.  Since this is
the defining property of a shock, we will call those solutions ``gauge
shocks''.  However, we must stress the fact that, in contrast to the
case of hydrodynamics, once a gauge shock forms one should make no
attempt to continue the solution any further since such a shock will
indicate that our coordinate system has broken down, and there is no
physical principle that can be used to extend the solution beyond this
point.  We will therefore not have shock waves as such
(i.e. propagating shocks), but just shock formation.

Gauge shocks were first studied in
References~\cite{Alcubierre97a,Alcubierre97b}, where it was found that
discontinuities in the lapse can easily develop starting from smooth
initial data in a wide variety of cases.  It was also shown how, in
some particular cases, one can even predict the exact time when a
gauge shock would form by just analyzing the initial data.

In~\cite{Alcubierre97a} a particular condition on the function
$f(\alpha)$ was found for which gauge shocks would not form:
\begin{equation}
1 \,-\, f \,-\, \alpha \, f' / 2 \,=\ 0 \;\; .
\label{eq:fcondition}
\end{equation}
This condition, however, was derived using a hyperbolic formulation of
the 3+1 evolution equations (the Bona-Masso formulation).  In the
following sections I will re-derive the condition making no
reference to the Einstein equations in any form.


\subsection{Linear degeneracy and shocks}
\label{sec:degeneracy}

Consider a system of equations of the form
\begin{equation}
\partial_t u_i \,+\, \partial_x F_i \,=\, q_i  \qquad i \in
\{1,...,N_u\} \; ,
\label{eq:usystem1}
\end{equation}
where $F_i$ and $q_i$ are arbitrary, possibly non-linear,
functions of the $u$'s but not their derivatives.  Notice that the 
system above can be written also as
\begin{equation}
\partial_t \, u_i \,+\, \sum_j M_{ij} \partial_x u_j
\,=\, q_i \qquad i \in
\{1,...,N_u\} \; ,
\label{eq:usystem2}
\end{equation}
with $M_{ij}= \partial F_i / \partial u_j$ the Jacobian matrix.

Let $\lambda_i$ be the eigenvalues of the Jacobian matrix
$M$. The system of equations is called ``hyperbolic'' if all the
$\lambda_i$ are real.  Further, the system is said to be ``strongly
hyperbolic'' if the matrix $M$ has a complete set of eigenvectors
${\bf e}_i$ .  Let us assume that this is the case, we then define the
eigenfields $w_i$ in the following way
\begin{equation}
{\bf u} \,=\, R \; {\bf w} \quad \Rightarrow \quad
{\bf w} \,=\, R^{-1} \, {\bf u} \;\;,
\end{equation}
where $R$ is the matrix of column eigenvectors ${\bf e}_i$.  One can
show that the matrix $R$ is such that
\begin{equation}
R M R^{-1} = \Lambda \; ,
\end{equation}
with $\Lambda={\rm diag}(\lambda_i)$.  The evolution equation for the
eigenfields $w_i$ then turns out to be
\begin{equation}
\partial_t w_i \,+\, \lambda_i \; \partial_x w_i \,=\, q'_i   \; ,
\end{equation}
with $q'_i$ a function of the $w's$ but not their derivatives. In terms
of the eigenfields the system transforms into a series of coupled
advection equations with characteristic speeds given by the
eigenvalues $\lambda_i$.

If a given eigenvalue $\lambda_i$ is independent of its corresponding
eigenfield $w_i$, then we say that the eigenfield is ``linearly
degenerate''~\cite{Lax73,Leveque92}, that is
\begin{equation} \frac{\partial \,
\lambda_i}{\partial \, w_i} \,=\, \sum_{j=1}^{N_u} \; \frac{\partial \,
\lambda_i}{\partial \, u_j} \, \frac{\partial \, u_j}{\partial \, w_i}
\,=\, \nabla_u \, \lambda_i \cdot {\bf e}_i \,=\, 0 \;\;.
\label{eq:degeneracy}
\end{equation}

Linear degeneracy is a sufficient condition for there not to be shocks
associated with a given eigenfield.  One can understand this
intuitively by noticing that linear degeneracy implies that the
characteristic lines do not change in response to changes in the
field propagating along them.


\subsection{Avoiding gauge shocks}
\label{sec:gaugeshocks}

In order to study the effects of our slicing condition without having
to worry about the evolution of the spacetime itself we will now
assume that we have a known background spacetime with metric $g_{\mu
\nu}$.  In that spacetime, we will consider some initial spatial
slice, and then construct a foliation according to our slicing
condition.  Let $T$ be the time function associated with our foliation
(that is, each spatial hypersurface will correspond to $T= {\rm
constant}$).  In Sec.~\ref{sec:foliation} above we saw that for the
Bona-Masso family of slicing conditions, the function $T$ will satisfy
the following foliation equation
\begin{equation}
\left[ g^{\mu \nu} + \left( 1 - \frac{1}{f(\alpha)} \right)
n^\mu n^\nu \right] \nabla_\mu \nabla_\nu T  = 0 \, ,
\label{eq:generalwave2}
\end{equation}
where $n^\mu$ is the unit normal vector to the hypersurfaces and where
$\nabla_\mu$ denotes covariant differentiation with respect to the
4-metric $g_{\mu \nu}$.  The unit normal vector $n^\mu$ can be easily
constructed from the time function $T$ in the following way
\begin{equation}
n_\mu = \frac{-\nabla_\mu T}{\left( - \nabla_\nu T \; 
\nabla^\nu T \right)^{1/2}} \; ,
\end{equation}
where the overall minus sign is there to guarantee that we have a
future pointing normal vector.

Let us now calculate the increment in $T$ if we move a proper distance
$d \tau$ along the normal direction
\begin{equation}
d T = \left( d \tau \; n^\mu \right)  \nabla_\mu T = d \tau \left(
- \nabla_\mu T \; \nabla^\mu T \right)^{1/2} \;.
\end{equation}
On the other hand, from the definition of $\alpha$ we have
\mbox{$d \tau = \alpha dT$}.  Comparing both expressions we find
\begin{equation}
\alpha = \left( - \nabla_\mu T \; \nabla^\mu T \right)^{-1/2} \; .
\end{equation}
The normal vector then takes the form:
\begin{equation}
n_\mu = -\alpha \nabla_\mu T \; .
\label{eq:unitnormal}
\end{equation}

Consider now a particular point $\mathcal{P}$ in spacetime.  To study
the evolution of $T$ close to that point we construct locally flat
coordinates $(t,x^i)$, so the metric close to $\mathcal{P}$ becomes
the flat metric $\eta_{\mu \nu}$ and the Christoffel symbols
vanish.  Equation~(\ref{eq:generalwave2}) then reduces to
\begin{equation}
\left[ \eta^{\mu \nu} - a
n^\mu n^\nu \right] \partial_\mu \partial_\nu T  = 0 \, ,
\label{eq:generalwave3}
\end{equation}
where we have defined $a:= 1/f - 1$.  Expanding this equation we find
\begin{eqnarray}
& - \left( 1 + a (n^0)^2 \right) \; \partial_t^2 T
+ \left( \delta^{ij} - a n^i n^j \right) \;
\partial_i \partial_j T \nonumber \\
& - 2 a n^0 n^i \; \partial_i \partial_t T = 0 \; .
\label{eq:generalwave4}
\end{eqnarray}

Let us now define $\Pi := \partial_t T$ and $\Psi_i := \partial_i T$.
Equation~(\ref{eq:generalwave4}) can then be transformed into the system
\begin{eqnarray}
\partial_t \Pi    &=& - \sum_i \frac{2 a n^0 n^i}{1 + a (n^0)^2}
\; \partial_i \Pi \nonumber \\
&& + \sum_{ij} \frac{\delta^{ij} - a n^i n^j}{1 + a (n^0)^2}
\; \partial_i \Psi_j \; ,
\label{eq:pidot1} \\
\partial_t \Psi_i &=& \partial_i \Pi \; .
\label{eq:psidot1}
\end{eqnarray}

In our locally flat coordinate system, the
contravariant components of the unit normal vector become
\begin{equation}
n^0 = + \alpha \Pi \; , \qquad
n^i = - \alpha \Psi_i \; ,
\end{equation}
and the lapse reduces to
\begin{equation}
\alpha = \left( \Pi^2 - {\bf \Psi}^2 \right)^{-1/2} \; ,
\end{equation}
with ${\bf \Psi}^2 = \sum_i \Psi_i^2$.  The
system~(\ref{eq:pidot1})-(\ref{eq:psidot1}) then takes the following
form
\begin{eqnarray}
\partial_t \Pi &=& \sum_i \frac{2 a \alpha^2 \Pi \Psi_i}
{1 + a \alpha^2 \Pi^2} \; \partial_i \Pi \nonumber \\
&+& \sum_{ij} \frac{\delta_{ij} - a \alpha^2 \Psi_i \Psi_j}
{1 + a \alpha^2 \Pi^2}
\; \partial_i \Psi_j \; ,
\label{eq:pidot2}\\
\partial_t \Psi_i &=& \partial_i \Pi \; .
\label{eq:psidot2}
\end{eqnarray}

In order to see if our system of equations is hyperbolic we consider
derivatives along a fixed spatial direction, say $x$, and neglect
derivatives along different directions.  It is evident that in this
case the variables $\Psi_q$, with $q \ne x$, can be considered as
fixed and we only need to analyze the sub-system $\{\Pi,\Psi_x\}$. The
Jacobian matrix $M$ for this reduced system becomes:
\begin{equation}
M = \left(
\begin{array}{cc}
\displaystyle \frac{2 a \alpha^2 \Pi \Psi_x}{1 + a \alpha^2 \Pi^2} &
\displaystyle \frac{1 - a \alpha^2 \Psi_x^2}{1 + a \alpha^2 \Pi^2}
\\ 
1 & 0
\end{array}
\right) \; .
\label{eq:jacobian}
\end{equation}

The eigenvalues of this matrix are easily found to be
\begin{eqnarray}
\lambda_\pm &=& \frac{1}{1 + a \alpha^2 \Pi^2} \;
\left\{ a \alpha^2 \Pi \Psi_x \right. \nonumber \\
&& \pm \left. \left[ 1 + a \alpha^2 
\left( \Pi^2 - \Psi_x^2 \right) \right]^{1/2} \right\} \; ,
\label{eq:eigenvalues}
\end{eqnarray}
with corresponding eigenvectors
\begin{equation}
{\bf e}_\pm = \left( \lambda_\pm , 1 \right) \; .
\label{eq:eigenvectors}
\end{equation}

Having found the eigenvalues and eigenvectors we can now ask if our
system of equations is hyperbolic.  Consider first the case $f=0$.  In
such a case one can show that the eigenvalues reduce to
\begin{equation}
\lambda_\pm = \Psi_x / \Pi \; ,
\end{equation}
that is, both eigenvalues are equal and the
eigenvectors~(\ref{eq:eigenvectors}) do not form a complete set, so
the system is only weakly hyperbolic.

In the case when $f<0$, one can use the fact that \mbox{$\Pi^2>{\bf
\Psi}^2 \geq \Psi_x^2$} (which will hold for spacelike slices) to show
that the term inside the square root in
equation~(\ref{eq:eigenvalues}) is always negative and the system is
not hyperbolic.

The $f>0$ case turns out to be more difficult to analyze.  If
\mbox{$0<f \leq 1$}, then it is not difficult to prove that the term
inside the square root is always positive.  This means that we have
two distinct real eigenvalues and a complete set of eigenvectors, and
the system is therefore strongly hyperbolic.  If $f>1$, on the other
hand, the term inside the square root can become negative for
sufficiently large $\Psi_q^2 := \sum_{i \neq x} \Psi_i^2$.  It would
then appear that in such a case we do not have a hyperbolic system.
However, the fact that hyperbolicity depends on the size of $\Psi_q^2$
indicates that this is only a coordinate problem.  Indeed, if we
reorient our spatial coordinates in a way such that $n^\mu$ only has
components along the $(t,x)$ directions, then $\Psi_q^2$ vanishes and
the eigenvalues become real.  The fact that for a different
orientation of the spatial coordinates we can have complex eigenvalues
is just an indication that for $f>1$ the characteristic cones can tilt
beyond the $(t,x)$ coordinate plane.  This is analogous to solving the
hydrodynamic equations using a supersonic reference frame: the
hydrodynamic equations become elliptic, but this is just a consequence
of choosing a bad coordinate system.  We then conclude that for $f>0$,
an orientation of the coordinates always exists such that our system
of equations is strongly hyperbolic.

\vspace{5mm}

Using the expression for the eigenvectors above, the condition for
linear degeneracy, Eq.~(\ref{eq:degeneracy}), takes the form
\begin{equation}
C_\pm := \lambda_\pm \frac{\partial \lambda_\pm}{\partial \Pi}
+ \frac{\partial \lambda_\pm}{\partial \Psi_x} = 0 \; .
\end{equation}

A straightforward calculation gives the following independent linear
combinations of the previous conditions
\begin{eqnarray}
&& C_+ + C_- = 0 =  \alpha^2 \Pi \left[ 2 a \left( 1 + a \right) + \alpha a'
\right] \nonumber \\
&& \left\{ \frac{\alpha^2 \Pi^2 \left( 1 + a + a \alpha^2 \Psi_q^2 \right)
+ 3 \alpha^2 \left( \Pi^2 - \Psi_q^2 \right) - 3}
{\left( 1 + a \alpha^2 \Pi^2 \right)^3} \right\} , \hspace{2em}
\end{eqnarray}
\begin{eqnarray}
&& C_+ - C_- = 0 = - \alpha^2 \Psi_x^2 \left[ 2 a \left( 1 + a \right)
+ \alpha a' \right] \nonumber \\
&& \left\{ \frac{3 \alpha^2 \Pi^2 \left( 1 + a + a \alpha^2 \Psi_q^2 \right)
+ \alpha^2 \left( \Pi^2 - \Psi_q^2 \right) - 1}
{\left( 1 + a \alpha^2 \Pi^2 \right)^3 \sqrt{1 + a \alpha^2 \Psi_q^2}}
\right\}  , \hspace{2em}
\end{eqnarray}
where $a':= da/d \alpha$.
For arbitrary $\alpha$, $\Pi$ and $\Psi_i$, the only way in which both these
conditions can hold is if we take
\begin{equation}
2 a \left( 1 + a \right) + \alpha a' = 0 \; ,
\end{equation}
which turns out to be equivalent to
\begin{equation}
1 - f - \alpha f' / 2 = 0 \; .
\end{equation}
This is precisely the condition~(\ref{eq:fcondition}) found in
reference~\cite{Alcubierre97a}.  The main difference between the
derivation above and the one of reference~\cite{Alcubierre97a} is that
in that reference the condition was derived using a hyperbolic
formulation of the Einstein equations (the Bona-Masso formulation),
while the new derivation is based purely on analyzing the slicing
condition and makes no reference to the Einstein equations in any
form, showing the generality of the condition.

\vspace{5mm}

As already shown in reference~\cite{Alcubierre97a},
condition~(\ref{eq:fcondition}) can be trivially solved to find
\begin{equation}
f = 1 + k/\alpha^2 \; ,
\label{eq:fsolution}
\end{equation}
with $k>0$ an arbitrary constant.  For $k=0$ we recover harmonic
slicing.  On the other hand, if $k \neq 0$ we see that for small
$\alpha$ we have $f \sim \alpha^{-2}$, so the results of
Sec.~\ref{sec:focusing} imply that the slicing will be strongly
singularity avoiding.  However, we can also see that in this case we
are in the regime for which the lapse can easily become negative.
This means that the solution~(\ref{eq:fsolution}) has a serious
drawback for any non-zero value of $k$, since it can allow the lapse to
become negative as it collapses toward zero.

In the next sections we will see how one can still obtain useful
slicing conditions by looking for approximate solutions to
condition~(\ref{eq:fcondition}).


\subsection{Zero order shock avoidance}
\label{sec:zeroorder}

In the previous section we found that if we want to guarantee that no
shocks will form, then we must choose the function $f(\alpha)$ in a
way that is incompatible with having a lapse that does not become
negative when it collapses toward zero (except in the specific case of
harmonic slicing).  Here we will relax our requirements and look for
approximate solutions of condition~(\ref{eq:fcondition}).

We start by assuming that the lapse is very close to $1$, that is
\begin{equation}
\alpha = 1 + \epsilon \; ,
\end{equation}
with $\epsilon \ll 1$.   Notice that the limit above applies to
situations that are close to flat space, but generally does not apply
to strong field regions (like the region close to a black hole) where
the lapse can be expected to be considerably smaller than 1.  However,
in such regions considerations about singularity avoidance are
probably more important.  Our aim is to find slicing conditions that
can avoid singularities in strong field regions, and at the same time
don't have a tendency to generate shocks in weak field regions.

We can now expand $f$ in terms of $\epsilon$ as
\begin{eqnarray}
f &=& a_0 + a_1 \epsilon + {\cal O}(\epsilon^2) \nonumber \\
  &=& a_0 + a_1 ( \alpha - 1 ) + {\cal O}(\epsilon^2) \; ,
\label{eq:firstorder}
\end{eqnarray}
and look for solutions to~(\ref{eq:fcondition}) to lowest order in
$\epsilon$.

Substituting~(\ref{eq:firstorder}) into~(\ref{eq:fcondition}) we find
\begin{equation}
1 - a_0 - a_1 / 2 + {\cal O}(\epsilon) = 0 \; .
\end{equation}
This means that if we want condition~(\ref{eq:fcondition}) to
be satisfied to zero order in $\epsilon$ we must have
\begin{equation}
a_1 = 2 \left( 1 - a_0 \right) \; ,
\end{equation}
which implies
\begin{eqnarray}
f &=& a_0 + 2 \left( 1 - a_0 \right) \epsilon + {\cal O}(\epsilon^2)
\nonumber \\
&=& \left( 3 a_0 - 2 \right) + 2 \left( 1 - a_0 \right) \alpha
+ {\cal O}(\epsilon^2) \; .
\label{eq:fsolution1}
\end{eqnarray}

We must remember that~(\ref{eq:fsolution1}) is just an expansion for
small $\epsilon$.  Any form of the function $f(\alpha)$ that has the
same expansion to first order in $\epsilon$ will also satisfy
condition~(\ref{eq:fcondition}) to zero order. One family of such
functions emerges if we ask for $f(\alpha)$ to have the following form
\begin{equation}
f = \frac{p_0}{1 + q_{1} \epsilon} \; .
\end{equation}

It is not difficult to show that for this to have an expansion of the
form~(\ref{eq:fsolution1}) we must ask for
\begin{equation}
p_0 = a_0 \; , \qquad q_1 = 2 \left( a_0 - 1 \right) / a_0 \; ,
\end{equation}
which implies
\begin{eqnarray}
f &=& \frac{a_0^2}{a_0 + 2 \left( a_0 - 1 \right) \epsilon} \nonumber \\
&=&  \frac{a_0^2}{\left( 2 - a_0 \right) + 2 \left( a_0 - 1 \right) \alpha}
\; .
\label{eq:frational1}
\end{eqnarray}

Notice that if we take $a_0=1$, we recover harmonic slicing.  But
there is one other case that is of special interest: For $a_0=2$ the
previous solution reduces to $f=2/\alpha$, which corresponds to a
member of the 1+log family.  The crucial observation here is that, as
already mentioned in Sec.~\ref{sec:relating}, this specific member of
the 1+log family is precisely the one that has been found empirically
to be very robust in black hole
simulations~\cite{Arbona99,Alcubierre01a,Alcubierre02a}.  The fact
that it is the only member of the 1+log family that satisfies
condition~(\ref{eq:fcondition}) even approximately means that one
should in fact expect it to be particularly well behaved.


\subsection{First order shock avoidance}
\label{sec:firstorder}

We can now go one order higher in $\epsilon$ to obtain other interesting
forms of $f$.  Taking now
\begin{eqnarray}
f &=& a_0 + a_1 \epsilon + a_2 \epsilon^2 + {\cal O}(\epsilon^3) \nonumber \\
  &=& a_0 + a_1 ( \alpha - 1 ) + a_2 ( \alpha - 1 )^2
+ {\cal O}(\epsilon^3) \; ,
\label{eq:secondorder}
\end{eqnarray}
we find, after substituting into condition~(\ref{eq:fcondition}), that
\begin{equation}
\left( 1 - a_0 - a_1 / 2 \right) - \left( 3 a_1 / 2 + a_2
\right) \epsilon + {\cal O}(\epsilon^2) = 0 \; .
\end{equation}
Condition~(\ref{eq:fcondition}) will be satisfied to
first order if we take
\begin{eqnarray}
a_1 &=& 2 \left( 1 - a_0 \right) \; , \\
a_2 &=& - 3 a_1 / 2 =
- 3 \left( 1 - a_0 \right) \; .
\end{eqnarray}
So the expansion of $f$ takes the form
\begin{eqnarray}
f &=& a_0 + 2 \left( 1 - a_0 \right) \epsilon - 3 \left( 1 - a_0 \right)
\epsilon^2 + {\cal O}(\epsilon^3) \nonumber \\
&=& \left( 6 a_0 - 5 \right) + 8 \left( 1 - a_0 \right) \alpha
 \nonumber \\
&& - 3 \left( 1 - a_0 \right) \alpha^2 + {\cal O}(\epsilon^3) \; .
\label{eq:fsolution2}
\end{eqnarray}

Just as before, we can now look for rational functions that have the
above expansion.  One such possibility is to ask for $f$ to have the
form
\begin{equation}
f = \frac{p_0}{1 + q_1 \epsilon + q_2 \epsilon^2} \; .
\end{equation}

In order for $f$ to have the expansion~(\ref{eq:fsolution2}) we must
take
\begin{eqnarray}
p_0 &=& a_0 \; , \\
q_1 &=& 2 \left( a_0 - 1 \right) / a_0 \; , \\
q_2 &=& \left( 1 - a_0 \right) \; \left[ 3 a_0
+ \left( 1 - a_0 \right) \right] / a_0^2 \; ,
\end{eqnarray}
which means that $f$ takes the final form
\begin{eqnarray}
f &=& \frac{a_0^3}{a_0^2 - 2 a_0 \left( 1 - a_0 \right) \epsilon
 + \left( 1 - a_0 \right) \left( 4 - a_0 \right) \epsilon^2} \nonumber \\
&=& \frac{a_0^3}{\left( 4 - 3 a_0 \right) + \alpha \left( 1 - a_0 \right)
\left[ \left( 4 - a_0 \right) \alpha - 8 \right] }
\; .
\end{eqnarray}
If we take $a_0=1$ we again recover harmonic slicing.  Another
interesting case is obtained by asking for
\begin{equation}
4 - 3 a_0 = 0 \quad \Rightarrow \quad a_0 = 4/3 \; ,
\end{equation}
since in that case we will have $f \sim \alpha^{-1}$ for small
$\alpha$, and as we have seen this implies good singularity
avoidance.  For such a choice the function $f$ reduces to
\begin{equation}
f = \frac{8}{3 \alpha \left( 3 - \alpha \right)} \; .
\label{eq:newf}
\end{equation}
For small $\alpha$, this form of $f$ behaves as a member of the 1+log
family.  Moreover, it satisfies condition~(\ref{eq:fcondition}) to
higher order than the usual choice $f=2/\alpha$.  One could worry
about the fact that for the choice~(\ref{eq:newf}) the function $f$
can become negative for $\alpha>3$.  However, such a situation is
unlikely to arise in practice since the initial lapse is always taken
to be at most 1 throughout the region of interest, and later evolution
usually makes it even smaller.  Because of these facts, the slicing
condition given by the choice~(\ref{eq:newf}) would seem to be a good
candidate for a robust slicing condition when evolving systems with
strong gravitational fields (including black holes).  Whether this is
true or not can only be settled by numerical experiments where this
form of $f$ is tested against more traditional choices.

Notice also that, whereas in the case when $f=2/\alpha$ the asymptotic
gauge speed in regions where $\alpha \sim 1$ is $\sqrt{2}
\sim 1.41$, in the case~(\ref{eq:newf}) the gauge speed in those
regions is only $\sqrt{4/3} \sim 1.15$, which is much closer to the
physical speed of light and might represent an extra advantage as
gauge effects will not propagate much faster than physical effects.


\section{Conclusion}
\label{sec:conclusion}

I have considered the Bona-Masso hyperbolic family of slicing
conditions and have studied under which circumstances such hyperbolic
slicings can avoid two different types of singularities: focusing
singularities, defined as those for which the spatial volume elements
vanish at a bounded rate, and gauge shocks, defined as coordinate
singularities for which the lapse becomes discontinuous as a
consequence of the crossing of the characteristic lines associated
with the propagation of the gauge.

In the case of focusing singularities, I have extended the analysis of
Bona {\em et. al.} and shown that, depending on the form that the
function $f(\alpha)$ defining the slicing takes in the limit of small
$\alpha$, one can have three different types of behavior: the lapse
vanishes before the spatial volume elements do (collapse of the lapse
and strong singularity avoidance), the lapse vanishes at the same time
as the spatial volume elements and the singularity is reached after an
infinite coordinate time (collapse of the lapse and marginal
singularity avoidance), or the lapse vanishes at the same time as the
volume elements but the singularity is still reached in a finite
coordinate time (collapse of the lapse but no singularity avoidance).
Harmonic slicing falls into the marginal singularity avoiding case,
whereas the more commonly used 1+log family falls into the strong
singularity avoiding case.

For the case of gauge shocks I have re-derived, in a way that is
independent of the Einstein equations, a condition on the function
$f(\alpha)$ found previously that avoids them.  This condition,
unfortunately, is severely restrictive and implies that the lapse can
easily become negative during evolutions.  I have therefore studied
different forms of the function $f(\alpha)$ that satisfy the condition
only approximately.  This study has shown that one specific member of
the 1+log family that has previously been found empirically to be
particularly robust, is in fact the only member of that family that
satisfies the condition for shock avoidance even to lowest order.  By
asking for the shock avoidance condition to be satisfied to higher
order, I have found a new form of the function $f(\alpha)$ that has
the potential of being even more robust than the 1+log slicings for
simulations of strongly gravitating systems.

As a final comment, it is important to mention that elliptic slicing
conditions such as maximal slicing can easily avoid both focusing
singularities (since the volume elements are not allowed to change)
and gauge shocks (since the speed of propagation is infinite), so they
should in principle be more robust that the slicings considered here.
Elliptic conditions, however, are considerably more computationally
expensive.  Not only that, but they are typically much more
restrictive.  For example, maximal slices might not always exist (they
typically do not exist in cosmological scenarios).  Finally, elliptic
equations require boundary conditions that might be difficult to
impose in some situations, in particular when one has internal
boundaries such as those associated with black hole excision.  It is
because of these reasons that one should consider alternatives to
elliptic gauge conditions, such as those studied here.


\acknowledgments

I wish to thank Bernd~Br\"ugmann, Dario~Nu\~nez, Ed~Seidel and Deirdre
Shoemaker for many useful comments.  I also thank the ``Caltech
Visitors Program for the Numerical Simulation of Gravitational Wave
Sources'' for their hospitality when completing the final version of
this manuscript.  This work was supported by CONACyT Mexico under the
repatriation program.


\bibliographystyle{prsty}
\bibliography{bibtex/references}


\end{document}